\documentclass{article}%
\usepackage{amsmath}
\usepackage{amsfonts}
\usepackage{amssymb}
\usepackage{graphicx}%
\newtheorem{theorem}{Theorem}

\newtheorem{remark}[theorem]{Remark}

\newcommand{\bpartial}{\mathop{\partial\kern -4pt\raisebox{.8pt}{$|$}}}
\newcommand{\bra}{\mathopen{[\kern-1.6pt[}}
\newcommand{\ket}{\mathclose{]\kern-1.5pt]}}
\newcommand{\bbra}{\mathopen{[\kern-2.2pt[\kern-2.3pt[}}
\newcommand{\bket}{\mathclose{]\kern-2.1pt]\kern-2.3pt]}}
\begin{document}

\title{The Dirac-Hestenes Equation for Spherical Symmetric Potentials in the
Spherical and Cartesian Gauges\footnote{Accepted for publication in 
International Journal of Modern Physics A.}}
\author{{\footnotesize Rold\~{a}o da Rocha}$^{1}${\footnotesize and Waldyr A.
Rodrigues Jr.}$^{2}${\footnotesize \ }\\$^{1}${\footnotesize Instituto de F\'{\i}sica Te\'orica }\\{\footnotesize Universidade Estadual Paulista }\\{\footnotesize Rua Pamplona 145, 
01405-900 S\~ao Paulo, SP, Brazil}\\{\footnotesize and}\\{\footnotesize Institute of Physics Gleb Wataghin }\\{\footnotesize IFGW, UNICAMP CP 6165}\\{\footnotesize 13083-970 Campinas SP, Brasil. e-mail: roldao@ifi.unicamp.br.}\\$^{2}${\footnotesize Institute of Mathematics, Statistics and Scientific
Computation }\\{\footnotesize IMECC, UNICAMP, CP 6065}\\{\footnotesize 13083-859 Campinas SP, Brazil. e-mail: walrod@ime.unicamp.br}}
\maketitle

\begin{abstract}
In this paper using the apparatus of the Clifford bundle formalism we show how
straightforwardly solve in Minkowski spacetime the Dirac-Hestenes
equation---\ which is an appropriate representative in the Clifford bundle of
differential forms of the usual Dirac equation--- \ by separation of variables
for the case of a potential having spherical symmetry in the
\textit{Cartesian} and \textit{spherical} gauges. We show that contrary to
what is expected at a first sight, the solution of the DHE in both gauges has
exactly the same mathematical difficulty.

\end{abstract}

\section{Introduction}

In this paper the Clifford bundle formalism is used in order to show how to
solve in Minkowski spacetime the Dirac-Hestenes equation (DHE)---\ which is an
appropriate representative in the Clifford bundle of differential forms of the
usual Dirac equation--- \ by separation of variables for the case of a
potential having spherical symmetry using the Cartesian and spherical
gauges\footnote{See below for the precise definition of these terms.
\par
{}}. Our main result is that contrary to what is expected at a first sight
\cite{cook}, the finding of solutions of the DHE in any one of the mentioned
gauges presents exactly the same mathematical difficulty.

We hope that our approach which uses intrinsic methods and no matrix
representations helps to clarify some misunderstandings appearing in the
literature relative to: (i) the meaning and nature of Dirac-Hestenes spinor
fields (DHSF), which are sections of the spin-Clifford bundle and their
representatives in the Clifford bundle of differential forms, and (ii) the
relation between the familiar Dirac equation (satisfied by covariant Dirac
spinor fields) and the DHE, and its different expressions in different (spin
coframe) gauges and in different coordinate charts. Section 2 present in
condensed form some necessary mathematical preliminaries, whose details may be
found in \cite{rod2004,moro, rodol}. In Section 2, we discuss how to obtain
solutions of the DHE in a given potential exhibiting spherical symmetry.
First, solutions are obtained in detail in Section 3.1 in the
\textit{shperical gaug}e. In Section 3.2 the DHE is presented in the
\textit{Cartesian gauge}. It appears, at first sight that the equation in the
spherical gauge is more complicated than the DHE (for the same problem) in the
Cartesian gauge. However, this is not the case. Indeed, we succeeded in
putting the equations in both guages in forms in which it becomes obvious that
their solutions are easily obtained in exactly the same way. In Section 4 we
present our conclusions.

\section{Preliminaries}

In this paper $\mathcal{M=}(M\simeq\mathbb{R}^{4},%
\mbox{\boldmath{$\eta$}}%
,D,\tau_{%
\mbox{\boldmath{$\eta$}}%
},\uparrow\mathbf{)}$ denotes Minkowski spacetime structure\footnote{Note that
$%
\mbox{\boldmath{$\eta$}}%
\in\sec T_{2}^{0}M$ is the Minkowski metric, $D$ is the Levi-Civita connection
of $%
\mbox{\boldmath{$\eta$}}%
$, $\tau_{%
\mbox{\boldmath{$\eta$}}%
}\in\sec%
{\displaystyle\bigwedge\nolimits^{4}}
T^{\ast}M$ defines a spacetime orientation and $\uparrow$ refers to a time
orientation. Also, $\mathtt{\eta}\in\sec T_{0}^{2}M$ denotes the metric of the
cotangent bundle. Details, may be found in\ \cite{rodol,rosha,sawu}.}. \ By
$F(M)$ we denote the (principal) bundle of frames and by $\mathbf{P}%
_{\mathrm{SO}_{1,3}^{e}}(M\mathbf{)}$ the orthonormal frame bundle.
$P_{\mathrm{SO}_{1,3}^{e}}(M)$ denotes the orthonormal coframe bundle. Since
Minkowski spacetime is a spin manifold there exists $\mathbf{P}_{\mathrm{Spin}%
_{1,3}^{e}}(M\mathbf{)}$ and $P_{\mathrm{Spin}_{1,3}^{e}}(M\mathbf{)}$ which
are respectively the spin frame bundle and the spin coframe bundle. To
continue we select the orthonormal coframe bundle and the spin coframe bundle
for our considerations. We recall that sections of $P_{\mathrm{SO}_{1,3}^{e}%
}(M\mathbf{)}$ are orthonormal coframes and that sections of $P_{\mathrm{Spin}%
_{1,3}^{e}}(M\mathbf{)}$ are also orthonormal coframes such that two coframes
differing by a $2\pi$ rotation are distinct and two coframes differing by a
$4\pi$ rotation are identified. We denote in what follows by
$s:P_{\mathrm{Spin}_{1,3}^{e}}(M\mathbf{)\rightarrow}P_{\mathrm{SO}_{1,3}^{e}%
}(M\mathbf{)}$ the fundamental mapping present in the definition of
$P_{\mathrm{Spin}_{1,3}^{e}}(M\mathbf{)}$ (see \cite{moro,rodol} for details).
Next we introduce the Clifford bundle of differential forms $\mathcal{C\ell
(}M,\mathtt{\eta})$ which is a vector bundle associated to\ $P_{\mathrm{Spin}%
_{1,3}^{e}}(M\mathbf{)}$ \ whose section are sums of nonhomogeneous
differential forms, which will be called Clifford fields. We recall that
\ $\mathcal{C\ell(}M,\mathtt{\eta})=P_{\mathrm{SO}_{1,3}^{e}}(M)\times
_{\mathrm{Ad}^{\prime}}\mathbb{R}_{1,3}$, where $\mathbb{R}_{1,3}%
\simeq\mathbb{H}(2)$ is the spacetime algebra. \ Details of the bundle
structure are as follows:

(i) Let $\mathbf{\pi}_{c}:\mathcal{C}\ell(M,\mathtt{\eta})\rightarrow M$ be
the canonical projection of $\mathcal{C}\ell(M,\mathtt{\eta})$ and let
$\{U_{\alpha}\}$ be an open covering of $M$. There are trivialization mappings
$\mathbf{\psi}_{i}:\mathbf{\pi}_{c}^{-1}(U_{i})\rightarrow U_{i}%
\times\mathbb{R}_{1,3}$ of the form $\mathbf{\psi}_{i}(p)=(\mathbf{\pi}%
_{c}(p),\psi_{i,x}(p))=(x,\psi_{i,x}(p))$. If $x\in U_{i}\cap U_{j}$ and
$p\in\mathbf{\pi}_{c}^{-1}(x)$, then
\begin{equation}
\psi_{i,x}(p)=h_{ij}(x)\psi_{j,x}(p)
\end{equation}
for $h_{ij}(x)\in\mathrm{Aut}(\mathbb{R}_{1,3})$, where $h_{ij}:U_{i}\cap
U_{j}\rightarrow\mathrm{Aut}(\mathbb{R}_{1,3})$ are the transition mappings of
$\mathcal{C}\ell(M,\mathtt{\eta})$. We recall that every automorphism of
$\mathbb{R}_{1,3}$ is \textit{inner. }Then,
\begin{equation}
h_{ij}(x)\psi_{j,x}(p)=g_{ij}(x)\psi_{i,x}(p)g_{ij}(x)^{-1} \label{4.4}%
\end{equation}
for some $g_{ij}(x)\in\mathbb{R}_{1,3}^{\star}$, the group of invertible
elements of $\mathbb{R}_{1,3}$.

(ii) As it is well known the group $\mathrm{SO}_{1,3}^{e}$ has a natural
extension in the Clifford algebra $\mathbb{R}_{1,3}$. Indeed we know that
$\mathbb{R}_{1,3}^{\star}$ (the group of invertible elements of $\mathbb{R}%
_{1,3}$) acts naturally on $\mathbb{R}_{1,3}$ as an algebra automorphism
through its adjoint representation. A set of \emph{lifts} of the transition
functions of $\mathcal{C}\ell(M,\mathtt{g})$ is a set of elements
$\{g_{ij}\}\subset$ $\mathbb{R}_{1,3}^{\star}$ such that if\footnote{Recall
that $\mathrm{Spin}_{1,3}^{e}=\{a\in\mathbb{R}_{1,3}^{0}:a\tilde{a}%
=1\}\simeq\mathrm{Sl}(2,\mathbb{C)}$ is the universal covering group of the
restricted Lorentz group $\mathrm{SO}_{1,3}^{e}$. Notice that $\mathbb{R}%
_{1,3}^{0}\simeq\mathbb{R}_{3,0}\simeq\mathbb{C}(2)$, the even subalgebra of
$\mathbb{R}_{1,3}$ is the Pauli algebra.}
\begin{align}
\mathrm{Ad}  &  :g\mapsto\mathrm{Ad}_{g},\nonumber\\
\mathrm{Ad}_{g}(a)  &  =gag^{-1},\forall a\in\mathbb{R}_{1,3}, \label{4.6bis}%
\end{align}
then $\mathrm{Ad}_{g_{ij}}=h_{ij}$ in all intersections.

(iii) Also $\sigma=\mathrm{Ad}|_{\mathrm{Spin}_{1,3}^{e}}$ defines a group
homeomorphism $\sigma:\mathrm{Spin}_{1,3}^{e}\rightarrow\mathrm{SO}_{1,3}^{e}$
which is onto with kernel $\mathbb{Z}_{2}$. We have that Ad$_{-1}=$ identity,
and so $\mathrm{Ad}:\mathrm{Spin}_{1,3}^{e}\rightarrow\mathrm{Aut}%
(\mathbb{R}_{1,3})$ descends to a representation of $\mathrm{SO}_{1,3}^{e}$.
Let us call $\mathrm{Ad}^{\prime}$ this representation, i.e., $\mathrm{Ad}%
^{\prime}:\mathrm{SO}_{1,3}^{e}\rightarrow\mathrm{Aut}(\mathbb{R}_{1,3})$.
Then we can write $\mathrm{Ad}_{\sigma(g)}^{\prime}a=\mathrm{Ad}_{g}%
a=gag^{-1}$.

(iv) It is clear then, that the structure group of the Clifford bundle
$\mathcal{C}\ell(M,\mathtt{\eta})$ is reducible from $\mathrm{Aut}%
(\mathbb{R}_{1,3})$ to $\mathrm{SO}_{1,3}^{e}$. Thus the transition maps of
the principal bundle of oriented Lorentz cotetrads $P_{\mathrm{SO}_{1,3}^{e}%
}(M)$ can be (through $\mathrm{Ad}^{\prime}$) taken as transition maps for the
Clifford bundle. We then have \cite{lawmi}
\begin{equation}
\mathcal{C}\ell(M,\mathtt{\eta})=P_{\mathrm{SO}_{1,3}^{e}}(M)\times
_{\mathrm{Ad}^{\prime}}\mathbb{R}_{1,3},
\end{equation}
i.e., the Clifford bundle is an associated vector bundle to the principal
bundle $P_{\mathrm{SO}_{1,3}^{e}}(M)$ of orthonormal Lorentz coframes.

\subsection{Clifford Fields}

\ Recall that $\mathcal{C}\!\ell(T_{x}^{\ast}M,\mathtt{\eta}_{x})$ is also a
vector space over $\mathbb{R}$ which is isomorphic to the exterior algebra
$\bigwedge T_{x}^{\ast}M$ of the cotangent space and $\bigwedge T_{x}^{\ast}M=%
{\displaystyle\bigoplus\nolimits_{k=0}^{4}}
\bigwedge{}^{k}T_{x}^{\ast}M$, where $\bigwedge^{k}T_{x}^{\ast}M$ is the
$\binom{4}{k}$-dimensional space of $k$-forms. There is a natural embedding
\ $\bigwedge T^{\ast}M\hookrightarrow$ $\mathcal{C}\ell(M,\mathtt{\eta})$
\cite{lawmi} and sections of $\mathcal{C}\!\ell(M,\mathtt{\eta})$---Clifford
fields ---can be represented as a sum of non-homogeneous differential forms.
Let $\{\mathbf{e}_{\mathbf{a}}\}\in\sec\mathbf{P}_{\mathrm{SO}_{1,3}^{e}}(M)$
(the orthonormal frame bundle) be a\textit{ }tetrad basis for $TU\subset TM$,
i.e., $g(\mathbf{e}_{\mathbf{a}},\mathbf{e}_{\mathbf{b}})=\eta_{\mathbf{ab}%
}=\mathrm{diag}(1,-1,-1,-1)$ and $(\mathbf{a,b}=0,1,2,3$). Moreover, let
$\{\mathbf{\varepsilon}^{\mathbf{a}}\}\in\sec P_{\mathrm{SO}_{1,3}^{e}}(M)$.
Then, \ for each $\mathbf{a}=0,1,2,3$, $\mathbf{\varepsilon}^{\mathbf{a}}%
\in\sec\bigwedge^{1}T^{\ast}M\hookrightarrow\sec\mathcal{C}\!\ell
(M,\mathtt{\eta})$, i.e., $\{\varepsilon^{\mathbf{b}}\}$ is the dual basis of
$\{\mathbf{e}_{\mathbf{a}}\}$. Finally, let $\{\mathbf{\varepsilon
}_{\mathbf{a}}\}$, $\mathbf{\varepsilon}_{\mathbf{a}}\in\sec\bigwedge
^{1}T^{\ast}M\hookrightarrow\sec\mathcal{C}\!\ell(M,\mathtt{\eta})$ be the
\textit{reciprocal basis} of $\{\mathbf{\varepsilon}^{\mathbf{b}}\}$, i.e.,
$\mathbf{\varepsilon}_{\mathbf{a}}\cdot\mathbf{\varepsilon}^{\mathbf{b}%
}=\delta_{\mathbf{a}}^{\mathbf{b}}.$

Recall also that the fundamental \emph{Clifford product }is generated by
\begin{equation}
\mathbf{\varepsilon}^{\mathbf{a}}\mathbf{\varepsilon}^{\mathbf{b}%
}+\mathbf{\varepsilon}^{\mathbf{b}}\mathbf{\varepsilon}^{\mathbf{a}}%
=2\eta^{\mathbf{ab}}. \label{00}%
\end{equation}
If $\mathcal{C}\in\sec\mathcal{C}\!\ell(M,\mathtt{\eta})$ is a Clifford field,
we have:%

\begin{equation}
\mathcal{C}=s+v_{\mathbf{i}}\mathbf{\varepsilon}^{\mathbf{i}}+\frac{1}%
{2!}b_{\mathbf{ij}}\mathbf{\varepsilon}^{\mathbf{i}}\mathbf{\varepsilon
}^{\mathbf{j}}+\frac{1}{3!}t_{\mathbf{ijk}}\mathbf{\varepsilon}^{\mathbf{i}%
}\mathbf{\varepsilon}^{\mathbf{j}}\mathbf{\varepsilon}^{\mathbf{k}%
}+p\mathbf{\varepsilon}^{\mathbf{5}}\;, \label{4.3bis}%
\end{equation}
where $\mathbf{\varepsilon}^{\mathbf{5}}=\mathbf{\varepsilon}^{\mathbf{0}%
}\mathbf{\varepsilon}^{\mathbf{1}}\mathbf{\varepsilon}^{\mathbf{2}%
}\mathbf{\varepsilon}^{\mathbf{3}}$ is the volume element and%
\begin{equation}
s,v_{\mathbf{i}},b_{\mathbf{ij}},t_{\mathbf{ijk}},p\in\sec%
{\displaystyle\bigwedge\nolimits^{0}}
T^{\ast}M\hookrightarrow\sec\mathcal{C}\!\ell(M,\mathtt{\eta}).
\end{equation}

Next we recall the crucial result \cite{moro,lawmi} that in a spin manifold we
have:
\begin{equation}
\mathcal{C}\ell(M,\mathtt{\eta})=P_{\mathrm{Spin}_{1,3}^{e}}(M)\times
_{\mathrm{Ad}}\mathbb{R}_{1,3}. \label{1new}%
\end{equation}

\subsection{Spinor Fields}

Spinor fields are sections of associated vector bundles to the principal
bundle of spinor coframes. The well known Dirac spinor fields are sections of
the bundle
\begin{equation}
S_{c}(M,\mathtt{\eta})=P_{\mathrm{Spin}_{1,3}^{e}}(M)\times_{\mu_{c}%
}\mathbb{C}^{4} \label{4.7}%
\end{equation}
$\mu_{c}$ the $D^{(1/2,0)}\oplus D^{(0,1/2)}$ representation of $\mathrm{Spin}%
_{1,3}^{e}\cong\mathrm{Sl}(2,\mathbb{C})$ in $\mathrm{End}(\mathbb{C}^{4})$
~\cite{choquet}.

Now, we introduce the \textit{left }spin-Clifford bundle, which is the
following associated vector bundle :
\begin{equation}
\mathcal{C}\ell_{\mathrm{Spin}_{1,3}^{e}}^{l}(M,\mathtt{\eta}%
)=P_{\mathrm{Spin}_{1,3}^{e}}(M)\times_{l}\mathbb{R}_{1,3}%
\end{equation}
where $l$ is the representation of $\mathrm{Spin}_{1,3}^{e}$ on $\mathbb{R}%
_{1,3}$ given by $l(a)x=ax$. Sections of $\mathcal{C}\ell_{\mathrm{Spin}%
_{1,3}^{e}}^{l}(M,\mathtt{\eta})$ are called left spin-Clifford fields.
$\mathcal{C}\ell_{\mathrm{Spin}_{1,3}^{e}}^{l}(M,\mathtt{\eta})$ is a
\ `principal $\mathbb{R}_{1,3}$-bundle', i.e., it admits a free action of
$\mathbb{R}_{1,3}$ on the right \cite{lawmi,moro,rodol}, which is denoted by
$R_{g}$, $g\in\mathbb{R}_{1,3}$. We shall need also to consider the
$\emph{right}$ real spin Clifford bundle for $M$ defined by%
\begin{equation}
\mathcal{C}\ell_{\mathrm{Spin}_{1,3}^{e}}^{r}(M,\mathtt{\eta}%
)=P_{\mathrm{Spin}_{1,3}^{e}}(M)\times_{r}\mathbb{R}_{1,3},
\end{equation}
where $r$ is the representation of $\mathrm{Spin}_{1,3}^{e}$ on $\mathbb{R}%
_{1,3}$ given by $r(a)x=xa$. Sections of $\mathcal{C}\ell_{\mathrm{Spin}%
_{1,3}^{e}}^{r}(M,\mathtt{\eta})$ are called right spin-Clifford fields. A
crucial result is the proposition proved in \cite{moro} that there is a
natural pairing%
\begin{equation}
\mathcal{C}\ell_{\mathrm{Spin}_{1,3}^{e}}^{l}(M,\mathtt{\eta})\times
\mathcal{C}\ell_{\mathrm{Spin}_{1,3}^{e}}^{r}(M,\mathtt{\eta})\rightarrow
\mathcal{C}\ell(M,\mathtt{\eta}).
\end{equation}
Such a proposition permits us to show that there is a \textit{well defined}
product of sections of $\mathcal{C}\ell_{\mathrm{Spin}_{1,3}^{e}}%
^{l}(M,\mathtt{\eta})$ by sections of $\mathcal{C}\ell_{\mathrm{Spin}%
_{1,3}^{e}}^{r}(M,\mathtt{\eta})$ \ and thus, a representation of any Clifford
field by a product of appropriate sections of $\mathcal{C}\ell_{\mathrm{Spin}%
_{1,3}^{e}}^{l}(M,\mathtt{\eta})$ by sections of $\mathcal{C}\ell
_{\mathrm{Spin}_{1,3}^{e}}^{r}(M,\mathtt{\eta})$.

The subbundle $I(M,\mathtt{\eta})$ of $\mathcal{C}\ell_{\mathrm{Spin}%
_{1,3}^{e}}^{l}(M,\mathtt{\eta})$ where the typical fiber is the ideal
$\mathbf{I}=\mathbb{R}_{1,3}\mathrm{e}$ (see below) is called the bundle of
left ideal algebraic spinor field (LIASF). Finally, we recall that there is a
\textit{natural} embedding $P_{\mathrm{Spin}_{1,3}^{e}}(M)\hookrightarrow
\mathcal{C}\ell_{\mathrm{Spin}_{1,3}^{e}}^{l}(M,\mathtt{\eta})$ which comes
from the embedding $\mathrm{Spin}_{1,3}^{e}\hookrightarrow\mathbb{R}_{1,3}%
^{0}$.

\subsection{Dirac-Hestenes Spinor Fields}

The importance of $\mathcal{C}\ell_{\mathrm{Spin}_{1,3}^{e}}^{l}%
(M,\mathtt{\eta})$ is that there are particular sections of this bundle that
are in one-to-one correspondence with \textit{Dirac} fields. This is seen as
follows. Let $\mathbf{E}^{\mu}$, $\mu=0,1,2,3$ be the canonical basis of
$\mathbb{R}^{1,3}\hookrightarrow\mathbb{R}_{1,3}$ which generates the algebra
$\mathbb{R}_{1,3}$. They satisfy the basic relation $\mathbf{E}^{\mu
}\mathbf{E}^{\nu}+\mathbf{E}^{\nu}\mathbf{E}^{\mu}=2\eta^{\mu\nu}$. We recall
that
\begin{equation}
\mathrm{e}\text{ }\mathbf{=}\frac{1}{2}(1+\mathbf{E}^{0})\in\mathbb{R}_{1,3}
\label{dh1}%
\end{equation}
is a primitive idempotent of $\mathbb{R}_{1,3}$ and
\begin{equation}
\mathbf{f}=\frac{1}{2}(1+\mathbf{E}^{0})\frac{1}{2}(1+\mathrm{i}\mathbf{E}%
^{2}\mathbf{E}^{1})\in\mathbb{C\otimes R}_{1,3} \label{dh2}%
\end{equation}
is a primitive idempotent of $\mathbb{C\otimes R}_{1,3}$. Now, let
$\mathbf{I}=\mathbb{R}_{1,3}\mathbf{e}$ and $\mathbf{I}_{\mathbb{C}%
}=\mathbb{C\otimes R}_{1,3}\mathbf{f}$ be respectively the minimal left ideals
of $\mathbb{R}_{1,3}$ and $\mathbb{C\otimes R}_{1,3}$ generated by
$\mathbf{e}$ and $\mathbf{f}$. Let $\mathbf{\phi=\phi}\mathrm{e}\in\mathbf{I}$
and $\mathbf{\Psi=\Psi f\in I}_{\mathbb{C}}$. Then, any $\mathbf{\phi\in I}$
can be written as
\begin{equation}
\mathbf{\phi=\psi}\mathrm{e} \label{dh3}%
\end{equation}
with $\mathbf{\psi}\in\mathbb{R}_{1,3}^{0}$. Analogously, any $\mathbf{\Psi\in
I}_{\mathbb{C}}$ can be written as
\begin{equation}
\mathbf{\Psi=\psi e}\frac{1}{2}(1+\mathrm{i}\mathbf{E}^{2}\mathbf{E}^{1}),
\label{dh4}%
\end{equation}
with $\mathbf{\psi}\in\mathbb{R}_{1,3}^{0}$.

Recall moreover that $\mathbb{C\otimes R}_{1,3}\simeq\mathbb{R}_{4,1}$
$\simeq\mathbb{C(}4)$, where $\mathbb{C(}4)$ is the algebra of the $4\times4$
complexes matrices. We can verify that
\begin{equation}
\left(
\begin{array}
[c]{cccc}%
1 & 0 & 0 & 0\\
0 & 0 & 0 & 0\\
0 & 0 & 0 & 0\\
0 & 0 & 0 & 0
\end{array}
\right)  \label{dh5}%
\end{equation}
is a primitive idempotent of $\mathbb{C(}4)$ which is a matrix representation
of $\mathbf{f}$. In that it can be proved that there is a bijection between
column spinors, i.e., elements of $\mathbb{C}^{4}$ (the complex $4$%
-dimensional vector space) and the elements of $\mathbf{I}_{\mathbb{C}}$.

Let $\mathbf{\Psi}\in\sec\mathcal{C}\ell_{\mathrm{Spin}_{1,3}^{e}}%
^{l}(M,\mathtt{\eta})$ be such that%
\begin{equation}
R_{\mathrm{e}}\mathbf{\Psi}=\mathbf{\Psi}\mathrm{e}=\mathbf{\Psi}%
,\mathrm{e}^{2}=\mathrm{e}=\mathbf{\frac{1}{2}(1+\mathbf{E}^{0})}\in
\mathbb{R}_{1,3}. \label{dh5'}%
\end{equation}
We define a Dirac-Hestenes Spinor field (DHSF) associated with $\Psi$ as an
\emph{even} section $\mathbf{\psi}$ of $\mathcal{C}\ell_{\mathrm{Spin}%
_{1,3}^{e}}^{l}(M,\mathtt{\eta})$ such that
\begin{equation}
\Psi=\mathbf{\psi}\mathrm{e}. \label{dh6}%
\end{equation}

\begin{remark}
An equivalent definition of a \emph{DHSF} is the following. Let $\mathbb{C}%
\ell_{\mathrm{Spin}_{1,3}^{e}}^{l}(M,\mathtt{\eta})=P_{\mathrm{Spin}_{1,3}%
^{e}}(M)\times_{l}\mathbb{C\otimes R}_{1,3}$ be the complex spin-Clifford
bundle. Let $\mathbf{\Psi}\in\sec\mathbb{C}\ell_{\mathrm{Spin}_{1,3}^{e}}%
^{l}(M,\mathtt{\eta})$ be such that
\begin{equation}
R_{\mathbf{f}}\mathbf{\Psi}=\mathbf{\Psi f=\Psi},\text{ }\mathbf{f}%
^{2}=\mathbf{f=\frac{1}{2}(1+\mathbf{E}^{0})\frac{1}{2}(1+\mathrm{i}%
\mathbf{E}^{2}\mathbf{E}^{1})}\in\mathbb{C}\mathbf{\mathbb{\otimes}}%
\mathbb{R}_{1,3}.
\end{equation}
Then, a \emph{DHSF} associated with $\mathbf{\Psi}$ is an even section
$\mathbf{\psi}$ of $\mathcal{C}\ell_{\mathrm{Spin}_{1,3}^{e}}^{l}%
(M,\mathtt{\eta})\hookrightarrow\mathbb{C}\ell_{\mathrm{Spin}_{1,3}^{e}}%
^{l}(M,\mathtt{\eta})$ such that
\begin{equation}
\mathbf{\Psi}=\mathbf{\psi f.} \label{dh7}%
\end{equation}

\end{remark}

In what follows, when we refer to a DHSF \ $\mathbf{\psi}$ \ we omit for
simplicity the wording associated with $\Phi$ (or $\Psi$). It is very
important to observe that a DHSF is not a sum of even multivector fields
although, under a local trivialization, $\mathbf{\psi}$ $\in\sec
\mathcal{C}\ell_{\mathrm{Spin}_{1,3}^{e}}^{l}(M,\mathtt{\eta})$ for each $x\in
M$ is mapped on an even element\footnote{Note that it is meaningful to speak
about even (or odd) elements in $\mathcal{C}\ell_{\mathrm{Spin}_{1,3}^{e}}%
^{l}(M)$ since $\mathrm{Spin}_{1,3}^{e}\subseteq\mathbb{R}_{1,3}^{0}.$
\par
{}} of \ $\mathbb{R}_{1,3}$. We emphasize that a DHSF is a particular section
of a spinor bundle, not of the Clifford bundle. However, and this is a very
important fact, any DHSF has representatives in the Clifford bundle. This
happens essentially because $P_{\mathrm{Spin}_{1,3}^{e}}(M)$ is trivial, a
fact that permits for each trivialization (i.e., choice of a spin coframe
$\Xi\in\sec P_{\mathrm{Spin}_{1,3}^{e}}(M)$ such that $s(\Xi
)=\{\mathbf{\varepsilon}^{\mathbf{a}}\}\in\sec P_{\mathrm{SO}_{1,3}^{e}}(M)$ )
to define a `unit section' for the right spin-Clifford bundle, i.e.,
$\mathbf{1}_{\Xi}^{r}\in\sec\mathcal{C}\ell_{\mathrm{Spin}_{1,3}^{e}}%
^{r}(M,\mathtt{\eta})$ such that for each Dirac-Hestenes spinor field
$\mathbf{\Psi}\in\sec\mathcal{C}\ell_{\mathrm{Spin}_{1,3}^{e}}^{l}%
(M,\mathtt{\eta})$ we have an \textit{even} Clifford field $\psi_{\Xi}\in
\sec\mathcal{C}\ell^{(0)}(M,\mathtt{\eta})\subset\sec\mathcal{C}%
\ell(M,\mathtt{\eta})$ such that%
\begin{equation}
\mathit{\psi}_{\Xi}=\mathbf{\Psi1}_{\Xi}^{r}. \label{definition of phi_Xi}%
\end{equation}
The field $\mathit{\psi}_{\Xi}$, which is a nonhomogeneous sum of even
differential forms (and which looks like a superfield) is said to be the
\textit{representative} of a Dirac-Hestenes spinor field (or of \ a Dirac
spinor field) in the Clifford bundle.

\subsection{Dirac and Dirac-Hestenes Equations}

Using $\mathit{\psi}_{\Xi}$ we can write a \textit{representative} of the
Dirac equation satysfied by a DHSF $\mathbf{\Psi}$ in interaction with an
electromagnetic field $A\in\sec%
{\displaystyle\bigwedge\nolimits^{1}}
T^{\ast}M\hookrightarrow\sec\mathcal{C}\ell(M,\mathtt{\eta})$ in the Clifford
bundle. First we recall that if $\mathbf{E}^{\mu}$, $\mu=0,1,2,3$ is the
canonical basis of $\mathbb{R}^{1,3}\hookrightarrow\mathbb{R}_{1,3}$ than the
Dirac equation \ for a DHSF is \cite{moro}%
\begin{equation}
{\mbox{\boldmath$\partial$}}^{s}\mathbf{\Psi E}^{\mathbf{21}}+m\mathbf{\Psi
E}^{\mathbf{0}}-qA\mathbf{\Psi}=0. \label{DE}%
\end{equation}
where ${\mbox{\boldmath$\partial$}}^{s}$ is the (spin) Dirac operator action
on sections of $\mathcal{C}\ell_{\mathrm{Spin}_{1,3}^{e}}(M,\mathtt{\eta})$.
We have in an arbitrary gauge $%
\mbox{\boldmath{$\Xi$}}%
$ with $s(%
\mbox{\boldmath{$\Xi$}}%
)=\{\mathbf{\varepsilon}^{\mathbf{a}}\}$ that%
\begin{equation}
{\mbox{\boldmath$\partial$}}^{s}\mathbf{\Psi=\varepsilon}^{\mathbf{a}%
}D_{\mathbf{e}_{\mathbf{a}}}^{s}\mathbf{\Psi=\varepsilon}^{\mathbf{a}%
}(\partial_{\mathbf{e}_{\mathbf{a}}}^{s}\mathbf{\Psi}+\frac{1}{2}\overset{%
\mbox{\boldmath{$\Xi$}}%
}{\omega}_{\mathbf{e}_{\mathbf{a}}}\mathbf{\Psi}) \label{DE1}%
\end{equation}
where $D_{\mathbf{e}_{\mathbf{a}}}^{s}$ is the spinor covariant derivative,
$\partial_{\mathbf{e}_{\mathbf{a}}}^{s}$ is the spin-Pfaff derivative (details
on $\partial_{\mathbf{e}_{\mathbf{a}}}^{s}$ which are not going to be used
anymore in this paper may be found in \cite{moro}) and $\overset{%
\mbox{\boldmath{$\Xi$}}%
}{\omega}_{\mathbf{e}_{\mathbf{a}}}$ is the $%
{\displaystyle\bigwedge\nolimits^{2}}
T^{\ast}M$ connection 1-form in the gauge $%
\mbox{\boldmath{$\Xi$}}%
$ evaluated at the vector field $\mathbf{e}_{\mathbf{a}}\in\sec TM$ .

The representative of the Dirac equation (Eq.\ref{DE}) in the Clifford bundle
in the gauge $%
\mbox{\boldmath{$\Xi$}}%
$ called the DHE is (taking into account that ${\mbox{\boldmath$\partial$}}%
^{s}$ is represented in $\mathcal{C}\ell(M,\mathtt{\eta})$ by the operator
${\mbox{\boldmath$\partial$}}^{(s)}\mathit{\psi}_{\Xi}%
={\mbox{\boldmath$\partial$}}\mathit{\psi}_{\Xi}\mathbf{\varepsilon
}^{\mathbf{21}}-\frac{1}{2}\mathbf{\varepsilon}^{\mathbf{a}}\mathit{\psi}%
_{\Xi}\overset{%
\mbox{\boldmath{$\Xi$}}%
}{\omega}_{\mathbf{e}_{\mathbf{a}}}$)

\begin{equation}
{\mbox{\boldmath$\partial$}}\mathit{\psi}_{\Xi}\mathbf{\varepsilon
}^{\mathbf{21}}-\frac{1}{2}\mathbf{\varepsilon}^{\mathbf{a}}\mathit{\psi}%
_{\Xi}\overset{%
\mbox{\boldmath{$\Xi$}}%
}{\omega}_{\mathbf{e}_{\mathbf{a}}}+m\mathit{\psi}_{\Xi}\mathbf{\varepsilon
}^{\mathbf{0}}-qA\mathit{\psi}_{\Xi}=0, \label{2new}%
\end{equation}
where
\begin{equation}
{\mbox{\boldmath$\partial$}}=\mathbf{\varepsilon}^{\mathbf{a}}D_{\mathbf{e}%
_{\mathbf{a}}} \label{3new}%
\end{equation}
is the Dirac operator acting on sections of the Clifford bundle. The action of
the covariant derivative of a Clifford field $\mathcal{C\in}\sec
\mathcal{C}\ell(M,\mathtt{\eta})$ is given by the notable formula (see, e.g.,
\cite{moro}),
\begin{equation}
D_{\mathbf{e}_{\mathbf{a}}}\mathcal{C}=\partial_{\mathbf{e}_{\mathbf{a}}%
}\mathcal{C}+\frac{1}{2}[\omega_{\mathbf{e}_{\mathbf{a}}},\mathcal{C}],
\label{4new}%
\end{equation}
where $\partial_{\mathbf{e}_{\mathbf{a}}}$is the Pfaff derivative of form
fields, i.e., taking into account Eq.(\ref{4.3bis}),
\begin{equation}
\partial_{\mathbf{e}_{\mathbf{a}}}\mathcal{C}=\mathbf{e}_{\mathbf{a}%
}(s)+\mathbf{e}_{\mathbf{a}}(v_{\mathbf{i}})\mathbf{\varepsilon}^{\mathbf{i}%
}+\frac{1}{2!}\mathbf{e}_{\mathbf{a}}(b_{\mathbf{ij}})\mathbf{\varepsilon
}^{\mathbf{i}}\mathbf{\varepsilon}^{\mathbf{j}}+\frac{1}{3!}\mathbf{e}%
_{\mathbf{a}}(t_{\mathbf{ijk}})\mathbf{\varepsilon}^{\mathbf{i}}%
\mathbf{\varepsilon}^{\mathbf{j}}\mathbf{\varepsilon}^{\mathbf{k}}%
+\mathbf{e}_{\mathbf{a}}(p)\mathbf{\varepsilon}^{\mathbf{5}}. \label{5new}%
\end{equation}

We need also to recall that the relation of the $%
{\displaystyle\bigwedge\nolimits^{2}}
T^{\ast}M$ connection 1-forms in two different gauges $%
\mbox{\boldmath{$\Xi$}}%
$ and $%
\mbox{\boldmath{$\Xi$}}%
^{\prime}$ related by $S\in\sec\mathrm{Spin}_{1,3}^{e}(M)\hookrightarrow
\mathcal{C}\ell(M,\mathtt{\eta})$is given by \cite{moro}%
\begin{equation}
\overset{%
\mbox{\boldmath{$\Xi$}}%
^{\prime}}{\omega}_{\mathbf{X}}=S\overset{%
\mbox{\boldmath{$\Xi$}}%
^{\prime}}{\omega}_{\mathbf{X}}S^{-1}+(D_{\mathbf{X}}S)S^{-1},
\label{connection}%
\end{equation}
where $\mathbf{X\in}\sec TM\mathbf{.}$

\section{ Spherical Symmetric Solutions of the DHE}

It is supposed that when the potential $A$ has spherical symmetry, as it is
the case, e.g., in a hydrogen atom, that it is mathematically more simple to
solve the Dirac equation or the DHE in the Cartesian gauge than in the
spherical gauge. As will be shown below the mathematical difficult involved in
solving the DHE in any one of these gauges is exactly the same one. To
proceed, we define precisely some terms. Let $\{x^{\mu}\}$ be global
coordinate functions for $M$ in Einstein-Lorentz coordinate gauge, i.e.,
$e_{0}=\partial/\partial x^{0}\in\sec TM$ is an inertial reference frame and
$\{x^{\mu}\}$ is a naturally adapted coordinate system to $e_{0}$ (nacs%
$\vert$%
$e_{o}$), the coordinate functions $x^{i}$, $i=1,2,3$ being the Cartesian
coordinate functions of the 3-dimensional rest space of $e_{0}$. Let
$\{x^{\prime0}=x^{0},x^{\prime i}\}$ be spherical coordinate functions
naturally adapted to $e_{0}$, i.e., $(x^{\prime1},x^{\prime2},x^{\prime
3})=(r,\theta,\varphi)$ are the usual spherical coordinate functions of the
3-dimensional rest space of $e_{0}$ relative to a given \textit{space} point
\cite{rosha,rodol}.

We have now, the following two sections\footnote{Note that $\{e_{\mu}\}$ is a
section of $F(M)$ which also belongs to $\mathbf{P}_{\mathrm{SO}_{1,3}^{e}%
}(M\mathbf{)}$.} of $\{e_{\mu}\},\{\bar{e}_{\mu}\}\in\sec$ $\mathbf{P}%
_{\mathrm{SO}_{1,3}^{e}}(M\mathbf{)}$:%
\begin{align}
e_{\mu}  &  =\partial/\partial x^{\mu},\nonumber\\
e_{0}^{\prime}  &  =\partial/\partial x^{0},\text{ }e_{1}^{\prime}%
=\frac{\partial}{\partial r},\text{ }e_{2}^{\prime}=\frac{1}{r}\frac{\partial
}{\partial\theta},\text{ }e_{3}^{\prime}=\frac{1}{r\sin\theta}\frac{\partial
}{\partial\varphi}. \label{6new}%
\end{align}
The corresponding dual frames are the sections $\{\gamma^{\mu}\}$ and
$\{\gamma^{\prime\mu}\}$ of $P_{\mathrm{SO}_{1,3}^{e}}(M\mathbf{)}$, with
\begin{align}
\gamma^{\mu}  &  =dx^{\mu},\nonumber\\
\gamma^{\prime0}  &  =dx^{0}\text{, }\gamma^{\prime1}=dr,\text{ }%
\gamma^{\prime2}=rd\theta,\text{ }\gamma^{\prime3}=r\sin\theta d\varphi.\text{
} \label{7new}%
\end{align}

Let $\Xi,\Xi^{\prime}$ be two sections of $P_{\mathrm{Spin}_{1,3}^{e}%
}(M\mathbf{)}$ such that
\begin{equation}
s(\Xi)=\{\gamma^{\mu}\}\text{, }s(\Xi^{\prime})=\{\gamma^{\prime\mu}\}.\text{
} \label{8new}%
\end{equation}

The spin coframes $\Xi,\Xi^{\prime}$ are called respectively Cartesian and the
spherical gauges. Recall that $\omega_{e_{\mu}}=0$, but some of the
$\omega_{e_{\mu}^{\prime}}$ are non null (see below). We introduce yet another
Cartesian gauge $\Xi_{o}$ and another spherical gauge $\Xi_{s}$ (which are
convenient for doing calculations) by
\begin{align}
s(\Xi_{o})  &  =\{\Gamma^{\mu}\},\nonumber\\
\Gamma^{\mu}  &  =U\gamma^{\mu}U^{-1}\text{, }U=\mathrm{e}^{\gamma^{23}%
\frac{\pi}{4}}, \label{9new}%
\end{align}
and%
\begin{align}
s(\Xi_{s})  &  =\{\vartheta^{\mu}\},\nonumber\\
\vartheta^{\mu}  &  =\Omega\Gamma^{\mu}\Omega^{-1}, \label{10new}%
\end{align}
where $\Omega\in\sec\mathrm{Spin}_{1,3}^{e}(M)\hookrightarrow\sec
\mathcal{C}\ell(M,\mathtt{\eta})$ is given by
\begin{equation}
\Omega=\exp(\gamma^{12}\frac{\varphi}{2})\exp(\gamma^{31}\frac{\theta}{2}).
\label{11new}%
\end{equation}

The dual basis of $\{\Gamma^{\mu}\}\in\sec P_{\mathrm{SO}_{1,3}^{e}}(M)$ is
$\{\mathbf{e}_{\mu}\}\in\sec\mathbf{P}_{\mathrm{SO}_{1,3}^{e}}(M)$ with%
\begin{equation}
\mathbf{e}_{0}=\partial/\partial x^{0},\text{ }\mathbf{e}_{1}=\frac{1}{r}%
\frac{\partial}{\partial\theta},\text{ }\mathbf{e}_{2}=-\frac{\partial
}{\partial r},\text{ }\mathbf{e}_{3}=\frac{1}{r\sin\theta}\frac{\partial
}{\partial\varphi}. \label{12new}%
\end{equation}

To simplify the writing of formulas we denote in what follows the
representatives of a DHSF satisfying the DHE in the gauges $\Xi,\Xi_{0}$ and
$\Xi_{s}$ by%
\begin{align}
\mathit{\psi}_{\Xi}  &  :=\mathit{\psi}_{c},\text{ }\mathit{\psi}_{\Xi_{o}%
}:=\mathit{\psi}_{o}=\mathit{\psi}_{c}U^{-1},\nonumber\\
\mathit{\psi}_{\Xi_{s}}  &  :=\mathit{\psi}_{s}=\mathit{\psi}_{o}\Omega^{-1}.
\label{13new}%
\end{align}

It is important for what follows to take into account that $\mathit{\psi}%
_{c},\mathit{\psi}_{o}$ and $\mathit{\psi}_{s}$ are even sections of the
Clifford bundle. Then, each Clifford field can be expressed in any arbitrary
coordinate chart of $M$, and as usual (sloppy notation) we denote a given
coordinate expression of a Clifford field by the same symbol.

\subsubsection{Spherical Gauge}

We now investigate the solution of the DHE for $A=V(r)\vartheta^{0}%
=V(r)\gamma^{0}$ in the spherical gauge $\Xi_{s}$. In this case, the DHE is
\begin{equation}
\vartheta^{\mu}\left(  \partial_{\mathbf{e}_{\mu}}\mathit{\psi}_{s}+\frac
{1}{2}\overset{s}{\omega}_{\mathbf{e}_{\mu}}\right)  \mathit{\psi}%
_{s}\vartheta^{13}+m\mathit{\psi}_{s}\vartheta^{0}-qA\mathit{\psi}_{s}=0,
\label{23new}%
\end{equation}
where $\overset{s}{\omega}_{\mathbf{e}_{\mu}}\in\sec%
{\displaystyle\bigwedge\nolimits^{2}}
T^{\ast}M\hookrightarrow\sec\mathcal{C}\ell(M,\mathtt{\eta})$ is given by
\cite{moro}%
\begin{equation}
\overset{s}{\omega}_{\mathbf{e}_{\mu}}=2(\partial_{\mathbf{e}_{\mu}}%
\Omega)\Omega^{-1}, \label{24new}%
\end{equation}
where $\Omega$ is given by Eq.(\ref{11new}).

At first (and eventually, second) sight Eq.(\ref{23new}) is more difficult to
solve than the corresponding equation in the Cartesian gauge (Eq.(\ref{14'new}%
) below) because the $\vartheta^{\mu}$ are variable covector fields and some
of the $\overset{s}{\omega}_{\mathbf{e}_{\mu}}\neq0$. However, let us analyze
the term $\vartheta^{\mu}(\partial_{\mathbf{e}_{\mu}}\Omega)\Omega
^{-1}\mathit{\psi}_{s}$. We have%
\begin{align}
\vartheta^{\mu}(\partial_{\mathbf{e}_{\mu}}\Omega)\Omega^{-1}\mathit{\psi
}_{s}  &  =\Omega\Gamma^{\mu}\Omega^{-1}(\partial_{\mathbf{e}_{\mu}}%
\Omega)\Omega^{-1}\mathit{\psi}_{s}\nonumber\\
&  =-\Omega\Gamma^{\mu}(\partial_{\mathbf{e}_{\mu}}\Omega^{-1})\Omega
\mathit{\psi}_{s}. \label{25new}%
\end{align}

Now, since%
\begin{align}
\Gamma^{0}(\partial_{\mathbf{e}_{0}}\Omega^{-1})\Omega\mathit{\psi}_{s}  &
=0,\nonumber\\
\Gamma^{1}(\partial_{\mathbf{e}_{1}}\Omega^{-1})\Omega\mathit{\psi}_{s}  &
=\frac{\gamma^{1}}{2r}\cot\theta\mathit{\psi}_{s}-\frac{\gamma^{3}}%
{2r}\mathit{\psi}_{s},\nonumber\\
\Gamma^{2}(\partial_{\mathbf{e}_{2}}\Omega^{-1})\Omega\mathit{\psi}_{s}  &
=0,\nonumber\\
\Gamma^{3}(\partial_{\mathbf{e}_{3}}\Omega^{-1})\Omega\mathit{\psi}_{s}  &
=-\frac{\gamma^{1}}{2r}\cot\theta\mathit{\psi}_{s}+\frac{\gamma^{3}}%
{2r}\mathit{\psi}_{s}, \label{20new}%
\end{align}
the term%
\begin{equation}
\Gamma^{\mu}(\partial_{\mathbf{e}_{\mu}}\Omega^{-1})\Omega\mathit{\psi}_{s}=0,
\label{20'new}%
\end{equation}
and Eq.(\ref{23new}) becomes%

\begin{equation}
\vartheta^{\mu}\partial_{\mathbf{e}_{\mu}}\mathit{\psi}_{s}\vartheta
^{13}+m\mathit{\psi}_{s}\vartheta^{0}-qA\mathit{\psi}_{s}=0. \label{26new}%
\end{equation}

Writing
\begin{equation}
\mathit{\psi}_{s}=\mathit{\psi}_{s1}(r,\theta)\mathrm{e}^{(n\varphi
-Et)\vartheta^{13}}, \label{27new}%
\end{equation}
where $n\in\mathbb{Z}$ we can separate Eq.(\ref{26new}), once T we recall that
the Pfaff derivatives $\partial_{\frac{\partial}{\partial r}}$ and
$\partial_{\frac{\partial}{\partial\theta}}$ , (which in the following we
write simply as $\frac{\partial}{\partial r}$ and $\frac{\partial}%
{\partial\theta}$) act only on the components of the Clifford fields (see
Eq.(\ref{5new})) We get a trivial equation in the $\varphi$ variable and the
following equation for $\mathit{\psi}_{s1},$%
\begin{equation}
(\vartheta_{30}\partial_{\frac{\partial}{\partial r}}+\vartheta_{10}%
\partial_{\frac{\partial}{\partial\theta}})\mathit{\psi}_{s1}\vartheta
_{13}+\frac{n\vartheta_{20}}{r\sin\theta}\mathit{\psi}_{s1}+(E-V)\mathit{\psi
}_{s1}=-m\vartheta^{0}\mathit{\psi}_{s1}\vartheta^{0}, \label{28new}%
\end{equation}
where $n\in\mathbb{Z}$.

Next we write
\begin{equation}
\psi_{s1}(r,\theta)=\vartheta^{12}\,\gimel(r)\vartheta^{13}\zeta
(\theta)+\gimel(r)\tilde{\zeta}(\theta), \label{ansz}%
\end{equation}
and get
\begin{equation}
\sin\theta\left(  \frac{d\zeta(\theta)}{d\theta}+\gamma^{13}\kappa\zeta
(\theta)\right)  -\lambda\tilde{\zeta}(\theta)=0
\end{equation}
for the angular component, where $\kappa\in\mathbb{R}$ is the separation
constant. Eq.(\ref{ansz}) has the general solution \cite{kruger}%

\begin{equation}
\varsigma_{p\lambda}(\theta)=b\sin^{|\lambda|}\theta\exp(\vartheta^{13}%
\theta/2[2\vartheta^{13}\left\vert \lambda\right\vert \sin\theta
C_{p-1}^{\left\vert \lambda\right\vert +1}(\cos\theta)+(p+2\left\vert
\lambda\right\vert )C_{p}^{\left\vert \lambda\right\vert }(\cos\theta)
\end{equation}
where $C_{p}^{a}$ are the Gegenbauer polynomials defined by
\begin{equation}
(p+1)C_{p+1}^{a}(z)-2(p+a)zc_{p}^{a}(z)+(p+2a-1)C_{p-1}^{a}(z)=0,\quad
C_{-1}^{a}(z)=0,\;\;C_{0}^{a}(z)=1,
\end{equation}
where $p\in\mathbb{N},\;a\in\mathbb{R}^{+}$ and
\begin{equation}
b=\frac{2^{|\lambda|}\Gamma(|\lambda|)}{4\pi}\sqrt{\frac{p!}{\Gamma
(p+2|\lambda|+1)}}.
\end{equation}

The radial equation is
\begin{equation}
-\vartheta^{3}\frac{d\;\gimel(r)}{dr}\vartheta^{13}+\left(  \vartheta^{1}%
\frac{\kappa}{r}+\vartheta^{0}(E-V)\right)  \,\gimel(r)+m\gimel(r)\vartheta
^{0}=0
\end{equation}
that can be decomposed writing
\begin{equation}
\gimel(r)=\gimel_{0}(r)-\vartheta^{23}\gimel(r),
\end{equation}
as
\begin{align}
\frac{d\;\gimel_{1}(r)}{dr}+\frac{\kappa}{r}\;\gimel_{1}(r)+(V-m-E)\,\gimel
_{0}(r)  &  =0,\label{v1}\\
\frac{d\;\gimel_{0}(r)}{dr}-\frac{\kappa}{r}\;\gimel_{0}(r)+(E-m-V)\,\gimel
_{1}(r)  &  =0. \label{v2}%
\end{align}
These are the well known radial equations for the Dirac equation solution
concerning the hydrogen atom \cite{rose}, whose solutions are well known.

\subsection{ Cartesian Gauge}

We now investigate how to solve the DHE for $A=V(r)\gamma^{0}$ in a Cartesian
gauge. First, the DHE in the gauges $\Xi$ and $\Xi_{0}$ are respectively%
\begin{align}
\gamma^{\mu}\mathbf{\partial}_{e_{\mu}}\mathit{\psi}_{c}\gamma^{21}%
+m\mathit{\psi}_{c}\gamma^{0}-qA\mathit{\psi}_{c}  &  =0,\label{14new}\\
\Gamma^{\mu}\mathbf{\partial}_{e_{\mu}}\mathit{\psi}_{o}\gamma^{13}%
+m\mathit{\psi}_{o}\gamma^{0}-qA\mathit{\psi}_{o}  &  =0. \label{14'new}%
\end{align}

Taking into account the (obvious) operator identity
\begin{equation}
\Gamma^{\mu}\mathbf{\partial}_{e_{\mu}}=\vartheta^{\mu}\partial_{\mathbf{e}%
_{\mu}} \label{15new}%
\end{equation}
we can write Eq.(\ref{14'new}) as%
\begin{equation}
\vartheta^{\mu}\partial_{\mathbf{e}_{\mu}}\mathit{\psi}_{o}\gamma
^{13}+m\mathit{\psi}_{o}\gamma^{0}-qA\mathit{\psi}_{o}=0, \label{16new}%
\end{equation}
or%

\begin{equation}
\Omega\Gamma^{\mu}\Omega^{-1}\partial_{\mathbf{e}_{\mu}}\mathit{\psi}%
_{o}\gamma^{13}+m\mathit{\psi}_{o}\gamma^{0}-qA\mathit{\psi}_{o}=0
\label{17new}%
\end{equation}
which, after introducing
\begin{equation}
\mathit{\psi}=\Omega^{-1}\mathit{\psi}_{o},\text{ }A^{\prime}=\Omega
^{-1}A\Omega\label{18new}%
\end{equation}
becomes,%
\begin{equation}
\Gamma^{\mu}\partial_{\mathbf{e}_{\mu}}\mathit{\psi}\gamma^{13}-\Gamma^{\mu
}(\partial_{\mathbf{e}_{\mu}}\Omega^{-1})\Omega\mathit{\psi}+m\mathit{\psi
}\gamma^{0}-qA^{\prime}\mathit{\psi}=0. \label{19new}%
\end{equation}
which\ taking into account that according to Eq.(\ref{20'new}) $\Gamma^{\mu
}(\partial_{\mathbf{e}_{\mu}}\Omega^{-1})\Omega\mathit{\psi}=0$\ can be easily
be solved by separation of variables by writing
\begin{equation}
\mathit{\psi}=\mathit{\psi}_{1}(r,\theta)\mathrm{e}^{(n\varphi-Et)\gamma^{13}%
}. \label{21new}%
\end{equation}
We have a trivial differential equation in the $\varphi$ variable and the
following equation for $\mathit{\psi}_{1}$,%

\begin{equation}
(\gamma_{30}\frac{\partial}{\partial r}+\gamma_{10}\frac{\partial}%
{\partial\theta})\mathit{\psi}_{1}\gamma_{13}+\frac{n\gamma_{20}}{r\sin\theta
}\mathit{\psi}_{1}+(E-V)\mathit{\psi}_{1}=-m\gamma^{0}\mathit{\psi}_{1}%
\gamma^{0}. \label{22new}%
\end{equation}
which can be solved in exactly the same way that Eq.(\ref{28new}) has been
solved once we take into account that the $\{\gamma^{\mu}\}$ and the
$\{\vartheta^{\mu}\}$ satisfy the same algebraic relations. We obviously get
the same spectrum, as it may be.

\begin{remark}
An equation like Eq.(\ref{22new}) has been used by Kr\"{u}ger \cite{kruger}
and also Daviau \cite{daviau}. However those authors arrive at that equation
using what to us seems to be a completely ad hoc argument (also used by
Hestenes and Lasenby, Doran and Gull \cite{ldg} ) which involves: (i) a
confusion between active local Lorentz transformations and transformations
relating the different expressions of the representatives of a DHSF in
different gauges and (ii) a supposedly change of the Dirac operator under an
active change of a Lorentz gauge transformation generate by $\Omega$. Both
assumptions are nonsequitur and produce misunderstandings. The concept of
active Lorentz gauge transformations of a \emph{DHSF} and the \emph{DHE} have
been discussed in a thoughtful way in \cite{rorova,roro}. Fortunately,
Eq.(\ref{22new}) is a fidedigne one, for otherwise the interesting results
found by Kr\"{u}ger and Daviau should be considered wrong.
\end{remark}

\section{Conclusions}

In this paper we showed how to solve (in Minkowski spacetime) the DHE \ by
separation of variables for the case of a potential having spherical symmetry
in two different ways, i.e., using the Cartesian and spherical gauges. We show
that contrary to what is expected at a first sight, the solution of the DHE in
any one of those gauges presents \textit{exactly} the same mathematical difficulty.

We also clarified some misunderstandings appearing in the literature related
to the meaning and nature of DHSF, the DHE and its different expressions in
different (spin coframe) gauges and way the use of different coordinate charts
does not implies change of gauge. We conjecture that "tricks" analogous to the
ones used in this paper can be used to solve with the same mathematical
difficulties the DHE \ with potentials exhibiting some others symmetries, both
in the Cartesian gauge and also in the gauge exhibiting the symmetry of the
potential. We will discuss this issue in another paper.


\begin{thebibliography}{99}                                                                                               %


\bibitem {choquet}{\small Choquet-Bruhat, Y., DeWitt-Morette, C., and
Dillard-Bleick, M., \textit{Analysis, Manifolds and Physics (revised
edition)}, North-Holland Publ. Co, Amsterdam, 1977.}

\bibitem {cook}{\small Cook, A. H., On Separable Solutions of Dirac's Equation
for the Electron, \textit{Proc. R. Soc. London A, Math. and Phys. Sci.}
\textbf{383}, 247-278 (1982).}

\bibitem {daviau}{\small Daviau, C., Sur une \'{E}quation d'onde Relativiste
et ses Solutions \`{a} Sym\'{e}trie Interne, \textit{Annales de la Fond. L. de
Broglie} \textbf{26}, 699-724 (2001).}

\bibitem {hestenes}{\small Hestenes,D.,} {\small \textit{Space-Time Algebra},
Gordon and Breach, New York 1966. }

\bibitem {kruger}{\small Kr\"{u}ger H., New Solutions of the Dirac Equation
for Central Fields, Hestenes D., and Weingartshofer, A. (eds.), \textit{The
Electron}, Kluwer Acad. Publ., Dordrecht, 1991.}

\bibitem {ldg}{\small Lasenby, A., Doran, C. and Gull, S., Gravity, Gauge
Theories and Geometric Algebras, \textit{Phil. Trans. R. Soc. London A}
\textbf{358}, 487-582 (1998)}.

\bibitem {lawmi}{\small Lawson, H. Blaine, Jr. and Michelson, M. L.,
\textit{Spin Geometry}, Princeton University Press, Princeton, 1989.}

\bibitem {moro}{\small Mosna R. A. and Rodrigues, W. A. Jr. , The bundles of
algebraic and Dirac-Hestenes spinor fields, \textit{J. Math. Phys.}
\textbf{45}, 2945-2988 (200}4). {\small [mat-ph/021233]}

\bibitem {rod2004}{\small Rodrigues, W. A. Jr., Algebraic and Dirac-Hestenes
spinors and spinor fields, \textit{J. Math. Phys} \textbf{45}, 2908-2945
(2004). [mat-ph/021230]}

\bibitem {rorova}{\small Rodrigues, W. A. Jr., Rocha, R. , and Vaz, J. Jr.,
Hidden Consequence of Active Lorentz Invariance, \emph{Int. J. Geom. Meth.
Mod. Phys.} \textbf{2}, 305-357 (2005) [\texttt{math-ph/0501064}].}

\bibitem {roro}{\small Rodrigues, W. A. Jr., Rocha, R., Diffeomorphism
Invariance and Local Lorentz Invariance, plenary lecture at ICCA7
[\texttt{math-ph/0510026].}}

\bibitem {rosha}{\small Rodrigues, W. A. Jr. and Sharif, M., Rotating Frames
in RT: Sagnac's Effect in SRT and other Related Issues, \textit{Found. Phys}.
\textbf{31}, 1767-1784 (2001). }

\bibitem {rodol}{\small Rodrigues, W. A. Jr. and Oliveira, E. Capelas,
T\textit{he Many Faces of Maxwell, Dirac and Einstein Equations}, RP56/05
IMECC-UNICAMP, [http://www.ime.unicamp.br/rel\_pesq/2005/rp56-05.html]}

\bibitem {rose}{\small Rose M. E., \textit{Relativistic Electron Theory}, John
Wiley, New York 1961.}

\bibitem {sawu}{\small Sachs, R. K. and Wu, H., \textit{General Relativiy for
Mathematicians}, Springer-Verlag, Berlin, 1977.}
\end{thebibliography}
\end{document}